\documentclass[nofootinbib,twocolumn]{revtex4}
\usepackage{graphicx}
\usepackage{amssymb}
\usepackage{amsmath}
\usepackage{bm}
\usepackage{hyperref}%for .pdf
\begin{document}

\title{Spherical collapse for unified dark matter models}

\author{Thiago R. P. Caram\^es}\email{trpcarames@gmail.com}\author{J\'ulio C.~Fabris}\email{fabris@pq.cnpq.br}\author{Hermano~E.~S.~Velten}\email{velten@pq.cnpq.br}

\affiliation{Universidade Federal do Esp\'{\i}rito Santo (UFES), Vit\'oria, ES - Brazil}

\begin{abstract}
We study the non-linear spherical ``top hat'' collapse for Chaplygin and viscous unified cosmologies. The term unified refers to models where dark energy and dark matter are replaced by one single component. For the generalized Chaplygin gas (GCG) we extend previous results of [R. A. A. Fernandes {\it et al}. Physical Review D 85, 083501 (2012)]. We discuss the differences at non-linear level between the GCG with $\alpha=0$ and the $\Lambda$CDM model. We show that both are indeed different. The bulk viscous model which differs from the GCG due to the existence of non-adiabatic perturbations is also studied. In this case, the clustering process is in general suppressed and the viable parameter space of the viscous model that accelerates the background expansion does not lead to collapsed structures. This result challenges the viability of unified viscous models. \\


\end{abstract}

\maketitle

\section{Introduction}

Although the $\Lambda$CDM remains being the simplest and the more efficient way to describe most of the astrophysical and cosmological data, many doubts concerning the nature of dark energy and dark matter still persist. The cold dark matter paradigm is a robust way to explain the large scale structure of the universe although its difficulties in explaining the cored density profile of galaxies and the small number of satellites around galaxies, the so called ``small scale problems''. However, it is crucial to emphasize that these inconsistencies come from the results of numerical simulations which still do not take into account all the physics envolved in the structure formation process. Moreover, the cosmological constant $\Lambda$ suffers from serious theoretical problems since its inferred value from the observations is around 120 orders of magnitude smaller than the typical value of the vacuum energy density predicted by particle physics. 

Alternative cosmologies incorporate new physics to the standard context either by replacing pieces of the $\Lambda$CDM model or by adding new ingredients to it. The unified scenario adopts the former suggestion. Its main idea is to replace the two unknown dark components, namely, dark matter and dark energy by one single fluid. As candidates for this scenario one can cite the GCG and the bulk viscous fluid which have both exotic equations of state. In some sense, the unification scenario can be seen as a modified dark matter dominated model, i.e., dark matter has some unusual equation of state that drives the universe to a late accelerated phase.

The formation of structures like galaxies and cluster depends on the dynamics of the baryonic matter which consists of no more than 5\% of the energy budget of the universe. At the same time, baryons follow the dark matter potential wells. The main challenge for unified cosmologies relies on the fact that the fluid that clusters to form ``dark matter halos'' also has to accelerate the expansion. Therefore, the clustering process of such models is a very sensitive probe for constructing viable cosmological scenarios. The models studied here will include baryons in the dynamics both at background and perturbative level.

A famous candidate for the unified scenario is the Chaplygin gas \cite{Kamenshchik1}, but its generalized form (the GCG) is widely used. The pressure $p_{gc}$ of the GCG is given by \cite{Kamenshchik, gorini, Bento, carturan}
\begin{equation}
p_{gc}=-\frac{A}{\rho_{gc}^{\alpha}}
\end{equation}
where $\rho$ is the energy density and $A$ and $\alpha$ are constants. Fixing $\alpha=1$ one has the original Chaplygin gas \cite{j3}. 

In the limit $\alpha=0$ it reduces to a scenario where dark matter has a negative constant pressure and the resulting dynamics is exactly the same as the $\Lambda$CDM model at background level as we will show below. However, it still not clear whether both models also display the same dynamics at perturbative level. 

Some works in the literature claim that the GCG gas with $\alpha=0$ and the $\Lambda$CDM model are also indistinguishable at first and non-linear order \cite{martins, Avelino, Kunz, Aviles}. Indeed, by choosing suitable first order variables there is a mathematical mapping between both models. However, a contrary opinion can be found in Refs. \cite{julioalpha0, juliano} where it is argued that the same definition for the matter density contrast used in the standard cosmology, which is also adopted by the galaxy surveys when obtaining the data, has to be adopted for alternative models like the GCG model. Therefore, in this case the mathematical mapping no longer exist. 

References \cite{Neven1, Neven2} first investigated the non-linear clustering of the original $\alpha=1$ Chaplygin gas. The extension to the generalized case was performed in \cite{Copeland}. The recent work by Fernandes {\it et al} \cite{Rui} extended in more detail the analysis for the GCG model with different values of $\alpha$ (see also \cite{popolo}). In this contribution we will revisit the equivalence between $\Lambda$CDM and the GCG model with $\alpha=0$. We calibrate our numerical code in order to match the results of Ref. \cite{Rui} but we also include in our analysis the $\Lambda$CDM case.

Such essence of unifing dark energy and dark matter effects into a single fluid is also captured by other models. The possibility of a late time expansion driven by the viscous mechanism has been first adopted in \cite{viscous1}. We also discuss the spherical collapse for this model. Assuming the Eckart approach for dissipative fluids \cite{Eckart}, its equation of state is given by
\begin{equation}
\Pi= -\xi u^{\gamma}_{; \gamma}
\end{equation}
where $u^{\gamma}$ is the $4-$velocity of the fluid and $\xi$ is the coefficient of bulk viscosity. Usually one adopts the form
\begin{equation}
\xi=\xi_0\left(\frac{\rho_v}{\rho_{v0}}\right)^{\nu},
\end{equation}
where $\xi_0$ is a constant and $\rho_v$ is the density of the bulk viscous fluid. The subscript $0$ denotes today's values. The constant parameter $\nu$ assumes in principle any real value. However, one has to keep in mind that transport coefficients deduced in kinetic theory depend on positive powers of the temperature of the fluid \cite{Chap}. Therefore, negative values for $\nu$ sounds unphysically from the thermodynamical point of view. 

In the FRW metric the bulk viscous pressure reduces to $\Pi =-3H\xi$. Due to the dependence of the pressure $\Pi$ with the expansion $H$, it is usually difficult to solve analitically for the density $\rho_v$ when fluids other than the viscous one, e.g. baryons, are present into the dynamics. Therefore we will obtain in this work only numerical results for the viscous model even at the background level.
 
\section{The background dynamics}

We are interested in expansions of the type

\begin{equation}
H^2(z)=H^2_0\left[ \Omega_{b0}(1+z)^3+\Omega_{unif}(z)\right],
\end{equation}
where $H_0$ is the Hubble constant today, $\Omega_{b0}$ is the today's density parameter for the baryonic matter and $\Omega_{unif}=\rho_{unif}/\rho_{crt}$ is the fractionary density parameter for the fluid that unifies the dark sector with $\rho_{crt}$ being the critical density. 

For the GCG we have
\begin{equation}
\Omega_{gc}(z)=\Omega_{gc0}\left\{\bar{A}+(1-\bar{A})(1+z)^{3(1+\alpha)}\right\}^{\frac{1}{1+\alpha}},
\end{equation}
where $\bar{A}=A/\rho_{gc0}^{1+\alpha}$ is a dimensionless parameter. Note that for $\alpha=0$ the expansion reduces to

\begin{equation}
H^2_{gc}(z)=H^2_0\left[\Omega_{b0}(1+z)^3+\Omega_{gc0}\left\{\bar{A}+(1-\bar{A})(1+z)^3\right\}\right]
\end{equation}
and the $\Lambda$CDM model is recovered with the identifications $\Omega_{gc0}\bar{A}=\Omega_{\Lambda}$ and $\Omega_{gc0}(1-\bar{A})=\Omega_{dm0}$.

For an one-fluid approximation, i.e., when one does not take into account other components into the dynamics, the evolution of the bulk viscous model and the GCG are exactly the same. The equivalence $\Omega_{v}(z)\equiv\Omega_{gc}(z)$ implies in the correspondence $\alpha=-(\nu +1/2)$ and $\bar{A}=3\xi_0 H_0 /\rho_{v0}$. Of course, this is an unrealistic case. If, at least, the baryons are included, the viscous pressure will indirectly depend on $\Omega_{b}$ since $\Pi \equiv \Pi(\xi,\Omega_v,\Omega_b)=-3\xi(\Omega_v)H(\Omega_v,\Omega_b)$. The evolution of $\Omega_v$ is therefore obtained from the numerical solution of 
\begin{equation}
(1+z)\frac{d\Omega_v}{dz}-3\Omega_v+\tilde{\xi}\left(\frac{\Omega_v}{\Omega_{v0}}\right)^{\nu}\left[\Omega_{v}+\Omega_{b0}(1+z)^3\right]^{1/2}=0
\end{equation}
where we have defined the dimensionless parameter $\tilde{\xi}=\frac{24\pi G \xi_0}{H_0}\left(\frac{3 H^2_0}{8\pi G}\right)^{\nu}$. The initial condition $\Omega_v(z=0)=\Omega_{v0}$ will be fixed to the same value as $\Omega_{gc0}=0.95$.

\section{Equations}\label{Sec:eq}

Our goal here is to obtain the perturbed equations for the evolution of an overdense spherical region collapsing in an expanding universe. 

We follow Refs. \cite{Rui, Abramo, Abramo2} and references therein. Let us first define basic quantities. For the collapsed region one can write
\begin{eqnarray}
\vec{v}_c= \vec{u}_0 + \vec{v}_p, \\
\rho_c=\rho\left(1+\delta\right) , \\
p_c=p + \delta p.
\end{eqnarray}

The velocity of the collapsed region $\vec{v}_c$ can be seen as the balance between the background expansion and the peculiar motion.

The effective expansion rate of the collapsed region is written as
\begin{equation}
h=H+\frac{\theta}{3a},
\end{equation}
where $\theta=\nabla \cdot v_p$ and $ v_p$ the peculiar velocity field.

For the collapssing region one has to assure energy conservation. Therefore, each component $i$ obeys a separate equation of the type

\begin{equation}\label{eqdeltaorig}
\dot{\delta_i}=-3H(c^2_{eff_i}-w_i)\delta_i-\left[1+w_i+(1+c^2_{eff_i})\delta_i\right]\frac{\theta}{a}
\end{equation}

where the energy density contrast is defined as

\begin{equation}\label{deltadef}
\delta_i = \left(\frac{\delta \rho}{\rho}\right)_i,
\end{equation}
and the effective speed of sound is computed following $c^2_{eff_i} = (\delta p / \delta \rho)_i$.

The dynamics of the perturbed region will be governed by the Raychaudhuri equation
\begin{equation}\label{eqthetaorig}
\dot{\theta}+H\theta+\frac{\theta^2}{3a}=-4\pi Ga \sum_i (\delta\rho_i + 3\delta p_i)\ .
\end{equation}

For the sake of comparison we write down below separately the system of equations for the GCG, the $\Lambda$CDM and the bulk viscous model.

\subsection{The GCG model}

The equations for the spherical collapse of the GCG have been deduced in details in \cite{Rui}. For the system baryons plus GCG they read

\begin{equation} \label{bgcg}
\dot{\delta}_b = -\left(1+\delta_b\right)\frac{\theta}{a},
\end{equation}
\begin{eqnarray} \label{mgcg}
\dot{\delta}_{gc} =-3H(c^2_{eff_{gc}}-w_{gc})\delta_{gc}\nonumber\\-\left[1+w_{gc}+(1+c^2_{eff_{gc}})\delta_{gc}\right]\frac{\theta}{a},
\end{eqnarray}
\begin{equation} \label{thetagcg}
\dot{\theta}+H\theta+\frac{\theta^2}{3a}=-4\pi Ga \left[\rho_b\delta_b+\rho_{gc}\delta_{gc}\left(1 + 3c^2_{eff_{gc}}\right)\right]\ .
\end{equation}

These equations use the definitions
\begin{equation}\label{wgcg}
w_{gc}= -\frac{\bar{A}}{\bar{A}+\left(1-\bar{A}a^{-3(1+\alpha)}\right)},
\end{equation}
\begin{equation}\label{csgcg}
c^2_{eff_{gc}}=w_{gc} \frac{(1+\delta_{gc})^{-\alpha}-1}{\delta_{gc}},
\end{equation}
which have been defined in \cite{Rui}.

\subsection{The $\Lambda$CDM model}
For the $\Lambda$CDM model both the baryonic and the dark matter component are pressureless fluids. 

\begin{equation} \label{blambda}
\dot{\delta}_b=-\left(1+\delta_b\right)\frac{\theta}{a}, 
\end{equation}
\begin{equation} 
\dot{\delta}_{dm}=-\left(1+\delta_{dm}\right)\frac{\theta}{a},
\end{equation}
\begin{equation} \label{thetalambda}
\dot{\theta}+H\theta+\frac{\theta^2}{3a}=-4\pi Ga (\rho_b\delta_b+\rho_{dm}\delta_{dm})\ .
\end{equation}

Note that baryons and dark matter obey to similar equations. This exemplifies the known fact that baryonic matter tracks the dark matter potential wells after the decoupling.

A crucial aspect for the develoment of this work is that the system (\ref{blambda})-(\ref{thetalambda}) can not be recovered by setting $\alpha=0$ into the GCG system (\ref{bgcg})-(\ref{csgcg}).

\subsection{The Bulk viscous model}

Let us now consider the bulk viscosity into the unified models described by the above set of equations. In general this is done by adding the bulk viscous pressure term $\Pi$ to the standard kinetic pressure $p_k$. However, for our viscous fluid we will use the approximation $p_k = 0$ so that the pressure of the unified viscous fluid is dominated by the nonadiabatic contribution $\Pi$. This is equivalent to

\begin{equation}
p= p_k+\Pi\rightarrow\Pi\ .
\end{equation}

Therefore, the speed of sound is properly modified following
\begin{equation}
\label{sound}
c^2_{eff_v}=\frac{\delta \Pi}{\delta\rho_v}=\frac{\frac{\xi \theta}{a}-3H\delta\xi}{\delta \rho_v}=\frac{w_v}{\delta_v}\left[\frac{\theta}{3Ha}+\left(1+\delta_v\right)^{\nu}-1\right]\ ,
\end{equation}
where $\omega_{v} \equiv \Pi/\rho_v$ is viscous equation of state parameter. Note that $c^2_{eff_v}$ depends on the potential of the perturbed velocity field which is a very particular feature of the bulk viscous model.

Therefore taking into account the introduction of the viscous pressure the dynamical equations for $\delta_i$ and $\theta$ are properly modified

\begin{equation}\label{bv}
\dot{\delta}_b = -\left(1+\delta_b\right)\frac{\theta}{a},
\end{equation}

\begin{eqnarray}\label{dmv}
\dot{\delta}_v=-3H\left(c^2_{eff_v}-w_v\right)\delta_v\nonumber\\-\left[1+w_v+(1+c^2_{eff_v})\delta_v\right]\frac{\theta}{a},
\end{eqnarray}

\begin{equation}\label{thetav}
\dot{\theta}+H\theta+\frac{\theta^2}{3a}=-4\pi Ga \left[\rho_b \delta_b  + \rho_v\delta_v \left(1+3c^2_{eff_v}\right)\right]\ .
\end{equation}

%We can alternatively express those equations by expliciting the dependence of $\delta_i$ and $\theta$ upon the redshift $z$, namely

%\begin{eqnarray} \label{eqdelta}
%H(1+z)\delta_i^{'}=3H\left(c^2_{eff_i}-w_i+\frac{\omega_v}{3H\delta_i}(1+z)\theta\right)\delta_i\nonumber\\+\left[1+w_i+(1+c^2_{eff_i})\delta_i+\frac{\omega_v \theta}{3 H}(1+z)\right]\theta (1+z)
%\end{eqnarray}

%\begin{eqnarray} \label{eqtheta}
%H(1+z)\theta^{'}=H\theta+\frac{\theta^2}{3}(1+z)\nonumber\\+\frac{3H^2}{2}\sum_i \left[\Omega_i \delta_i\left(1+3c^2_{eff_i}\right)\left(1+z\right)^{-1}+\frac{\Omega_i\omega_v}{H}\theta \right]\ ,
%\end{eqnarray}
%where primes denote derivatives with respect to $z$.

\section{Results}

\subsection{Generalized Chaplygin gas {\it versus} $\Lambda$CDM}

We extend now the results of Ref. \cite{Rui} which did not compare the spherical collapse for the GCG with $\alpha=0$ with the dark matter component of the $\Lambda$CDM model. This analysis is very important because many studies in the literature argue that both models are identical also at perturbative level. Indeed, as discussed in \cite{martins, Avelino, Kunz, Aviles} the supposed equivalence between these models, known as ``dark degeneracy'', should also be observed at first order perturbative stage. They have shown however that this degeneracy survives at linear regime if the same initial conditions for $\delta$ and $\theta$ are assumed. 

We wonder if a dark matter dominated model in which its equation of state is a negative and constant pressure is exactly the same as the $\Lambda$CDM, 
then why it does not become the standard cosmological model?
 
In fact, if dark matter only leads to the same universe as the $\Lambda$CDM model, 
this would represent the final solution of the dark energy paradigm. Of course, the answer is not so simple and indeed there exist a difference between the models as we will show below.

The study of the non-linear clustering can break the degeneracy. If we trace the behavior of the dark matter perturbations it is possible to show that this component behaves differently in both models.

Following \cite{Rui} we solve numerically the system of equations (\ref{bgcg}) - (\ref{thetagcg}) with initial conditions $\delta_{gc}(z=1000)=3.5\times10^{-3}$, $\delta_b(z=1000)=10^{-5}$ and $\theta (z=1000)=0$. For the background we fix $H_0 =72Km/s/Mpc$, $\Omega_{gc0}=0.95$ and $\Omega_{b0}=0.05$. Besides, our analysis for the GCG always uses $\bar{A}=0.75$.  

A proper comparison of the GCG's results with the $\Lambda$CDM model occurs if we set $\Omega_{\Lambda}=\Omega_{gc0}\bar{A}=0.95\times0.75$=0.7125 and $\Omega_{dm0}=0.2375$ with the same values for $\Omega_{b0}$ and $H_0$. Then, this assures that the background evolution for the case $\alpha=0$ and the $\Lambda$CDM is exactly the same.

Fig. 1 shows the density perturbation growth of the dark component for both models. We plot the results for the GCG gas ($\delta_{gc}$) with values $\alpha=0$ (dashed black) and $\alpha=0.1$ (solid black). Our results match the results of \cite{Rui} and we have confirmed the consistence of our numerical code for all other values of $\alpha$. The dark matter density contrast $\delta_{dm}$ of a flat $\Lambda$CDM cosmology with $\Omega_{\Lambda}=0.7125$ is plotted in the solid red line of this same figure. As one can see, the behavior of the ``dark'' component in each model is indeed different and consequently, although very similar, the models are not the same. This conclusion is supported by a full analysis of the CMB power spectrum. To quote Amendola {\it et al}, ``We note how the $\Lambda$CDM curve and $\alpha=0$ are very close, but not identical because of the different perturbation sectors.'' \cite{carturan}. 

The agreement would exist if we had adopted for the GCG a different definition of the density contrast
\begin{equation}
\delta^{\star}_{gc}=\frac{\delta\rho_{ch}}{\rho_{gc}}(1+w_{gc})
\end{equation}
rather than the one defined in (\ref{deltadef}). However, the products of galaxy surveys are observables like the growth index, $\sigma_8$ and the matter power spectrum which adopt the definition (\ref{deltadef}). Therefore, this is the variable (\ref{deltadef}) that one has to assume when comparing the perturbative behavior of distinct cosmological models.

For the same cases studied in Fig. 1 we can plot the evolution of the baryonic component $\delta_b$, see Fig. 2, and the expansion rate $h$ of the collapsed region in Fig. 3. For both quantities there exist a perfect agreement between the GCG with $\alpha=0$ and the $\Lambda$CDM model.  

\begin{figure}\label{fig1}
\begin{center}
\includegraphics[width=0.43\textwidth]{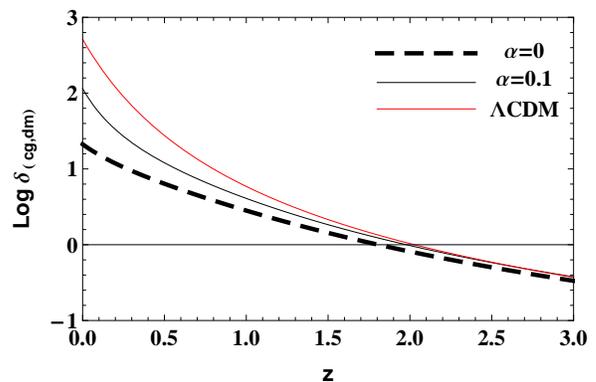}
%\label{fig1}	
\caption{Dark matter perturbation growth as function of the redshift. The black lines show $\delta_{gc}$ for $\alpha=0$ (dashed) and $\alpha=0.1$ (solid) fixing $\bar{A}=0.75$. Both agree with Ref. \cite{Rui}. Red line (upper one) shows $\delta_{dm}$ for the $\Lambda$CDM model.  }
\end{center}
\end{figure}

\begin{figure}\label{fig2}
\begin{center}
\includegraphics[width=0.43\textwidth]{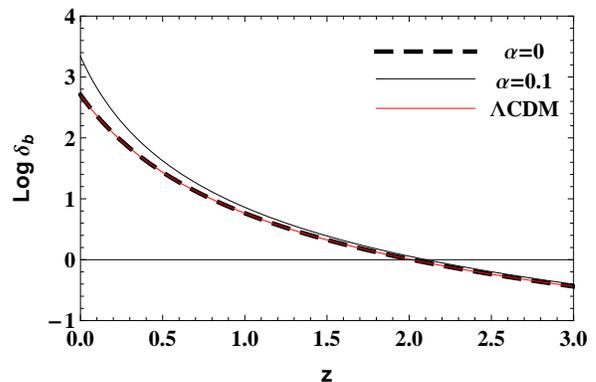}
%\label{fig1}	
\caption{Perturbation growth of baryons as a function of the redshift. The black lines show $\delta_{b}$ for GCG cosmologies with $\alpha=0$ (dashed) and $\alpha=0.1$ fixing $\bar{A}=0.75$. Both agree with Ref. \cite{Rui}. Red line shows $\delta_{b}$ for the $\Lambda$CDM model and agrees exactly with the case $\alpha=0$.  }
\end{center}
\end{figure}

\begin{figure}\label{fig3}
\begin{center}
\includegraphics[width=0.43\textwidth]{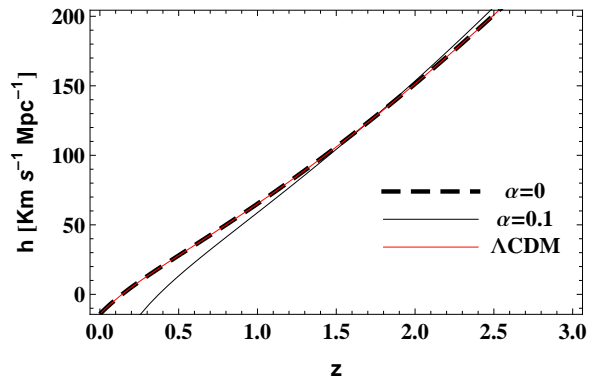}
%\label{fig1}	
\caption{Evolution of the expansion rate of the collapsed region. The black lines show $h$ for GCG cosmologies with $\alpha=0$ (dashed) and $\alpha=0.1$ fixing $\bar{A}=0.75$. Both agree with Ref. \cite{Rui}. Red line shows $h$ for the $\Lambda$CDM model and agrees exactly with the case $\alpha=0$.   }
\end{center}
\end{figure}

\subsection{Bulk viscous fluid}

For viscosities in the range $2 < \tilde{\xi} < 2.5$ the unified bulk viscous model explains the late phase of accelerated expansion and therefore it produces a viable background cosmology \cite{winfried, dominik, VeltenHz, MB}

We solve numerically the system of equations (\ref{bv})-(\ref{thetav}), with initial conditions $\delta_{v}(z=1000)=3.5\times10^{-3}$, $\delta_b(z=1000)=10^{-5}$ and $\theta (z=1000)=0$. For the background we fix $H_0 =72Km/s/Mpc$, $\Omega_{v0}=0.95$ and $\Omega_{b0}=0.05$. 

Let us start fixing $\nu=0$, i.e., the coefficient of bulk viscosity is a constant and its magnitude is given by $\tilde{\xi}$. We see in Figs. 4, 5 and 6 the evolution of the density contrast for the dark matter, baryons and the expansion of the collapsed, respectively. The results of Figs. 1, 2 and 3 are shown together in order to allow an appropriate comparison with the results of the GCG and the $\Lambda$CDM model. The bulk viscous fluid is displayed in dashed blue lines for values $\tilde{\xi}= 1$ (top line), $\tilde{\xi}=1.25$ (middle) and $\tilde{\xi}=1.5$ (bottom line). 

In Fig. 4 we see that dark structures are severely suppressed for the value $\tilde{\xi}=1.5$. Note that in this case the background expansion is not accelerated \cite{winfried, dominik, VeltenHz, MB}. The dissipative mechanism present in the bulk viscous fluid avoids the formation of viscous dark matter halos in the unified scenario. A similar problem occurs with the CMB data \cite{Barrow}.

In order to verify the consistence of this result we also plot the same as Figs. 4, 5 and 6 but for a coefficient $\nu=1/4$ in Figs. 7, 8 and 9. The dashed blue lines now correspond to the values $\tilde{\xi}=0.3$ (top line), $\tilde{\xi}=0.5$ (middle) and $\tilde{\xi}=0.7$ (bottom line). 

For the case $\nu=1/2$ we produce similar plots in Figs. 10, 11 and 12. The dashed blue lines now correspond to the values $\tilde{\xi}=0.3$ (top line), $\tilde{\xi}=0.5$ (middle) and $\tilde{\xi}=0.7$ (bottom line). The larger the coefficient $\nu$ the smaller the viscosity value $\tilde{\xi}$.

Remember that in kinetic theory the transport coefficients are calculated as a function of positive powers of the temperature of the fluid $\xi \equiv \xi(T)$ \cite{Chap}. Consequently negative values for $\nu$ have to be avoided. The range $0\leq\nu\leq 0.5$ covers the effective region of the bulk viscous parameter space.

\begin{figure}\label{fig4}
\begin{center}
\includegraphics[width=0.43\textwidth]{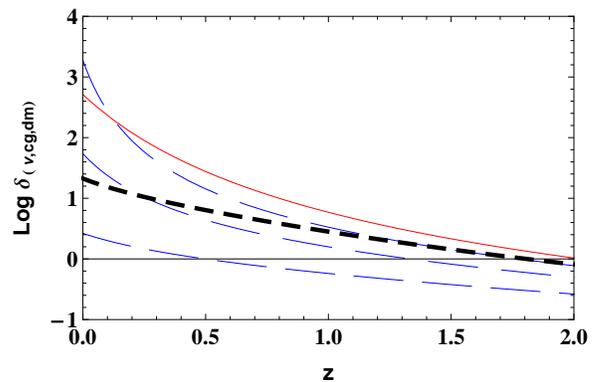}
%\label{fig1}	
\caption{Dark matter perturbation growth as function of the redshift. The same as Fig.1 but now we add the perturbation in the bulk viscous fluid $\delta_v$ (dashed blue lines) with, from top to bottom, $\tilde{\xi}=1, 1.25$ and $1.5$ fixing $\nu=0$. }
\end{center}
\end{figure}

\begin{figure}\label{fig5}
\begin{center}
\includegraphics[width=0.43\textwidth]{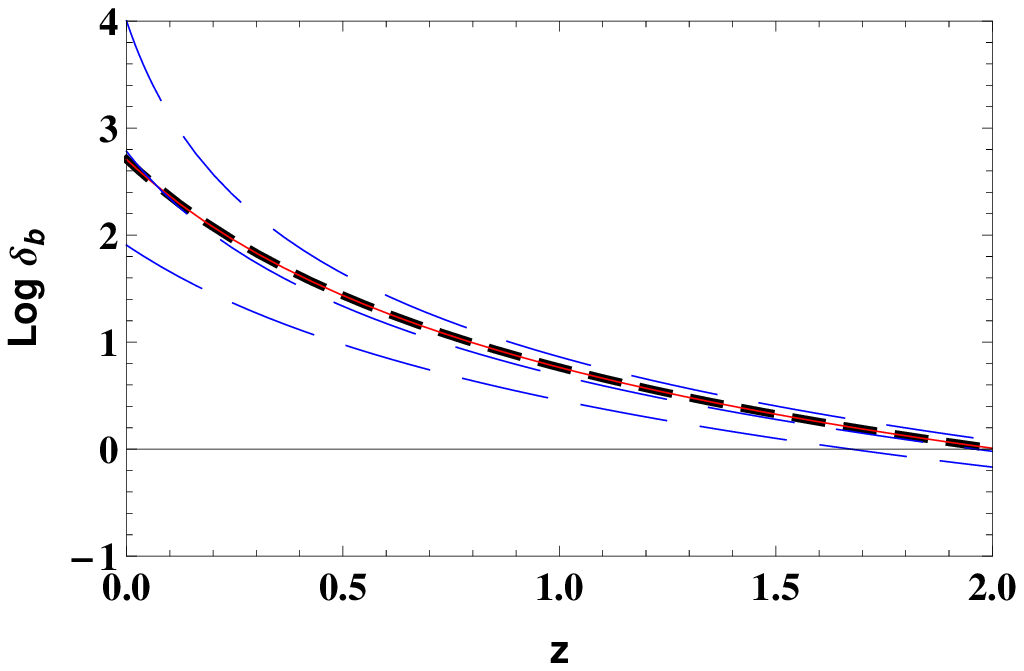}
%\label{fig1}	
\caption{Perturbation growth of baryons as a function of the redshift. The same as Fig.2 but now we add the baryonic perturbation in the bulk viscous fluid model (dashed blue lines) with, from top to bottom, $\tilde{\xi}=1, 1.25$ and $1.5$, fixing $\nu=0$.}
\end{center}
\end{figure}

\begin{figure}\label{fig6}
\begin{center}
\includegraphics[width=0.43\textwidth]{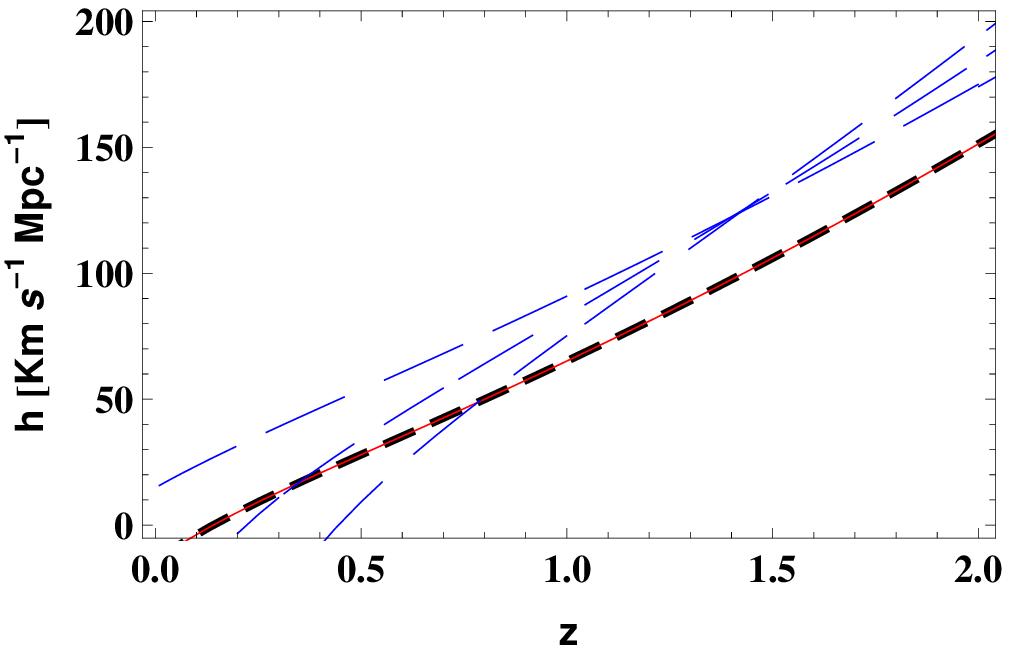}
%\label{fig1}	
\caption{Evolution of the expansion rate of the collapsed region. The same as Fig.3 but now we add the result for the bulk viscous fluid (dashed blue lines) with, from top to bottom, $\tilde{\xi}=1, 1.25$ and $1.5$ fixing $\nu=0$. }
\end{center}
\end{figure}

\begin{figure}\label{fig7}
\begin{center}
\includegraphics[width=0.43\textwidth]{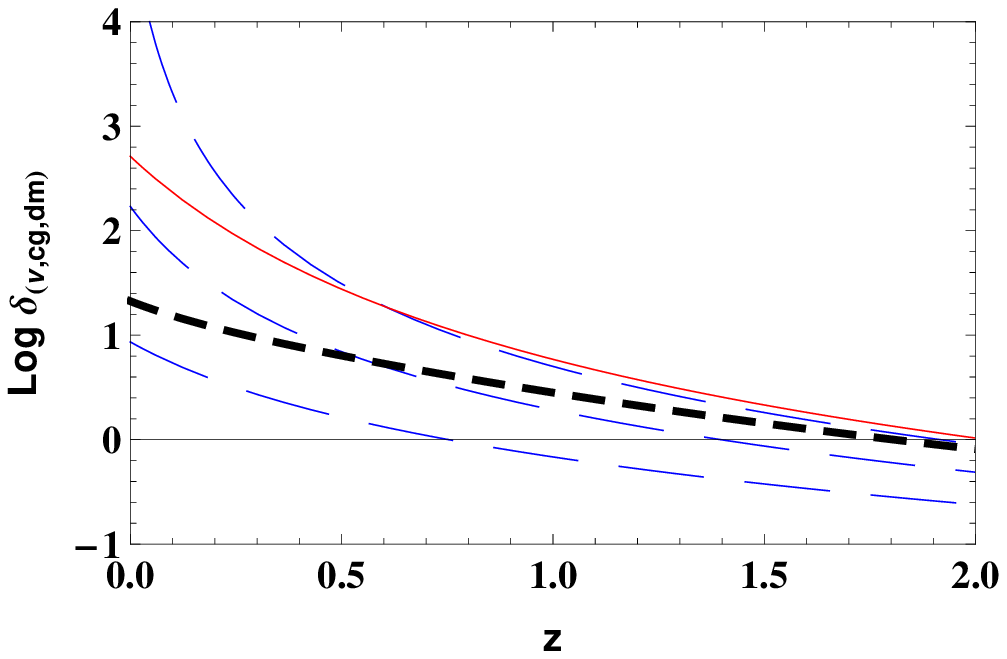}
%\label{fig1}	
\caption{Dark matter perturbation growth as function of the redshift. The same as Fig.1 but now we add the perturbation in the bulk viscous fluid $\delta_v$ (dashed blue lines) with, from top to bottom, $\tilde{\xi}=0.3, 0.5$ and $0.7$ fixing $\nu=1/4$. }
\end{center}
\end{figure}

\begin{figure}\label{fig8}
\begin{center}
\includegraphics[width=0.43\textwidth]{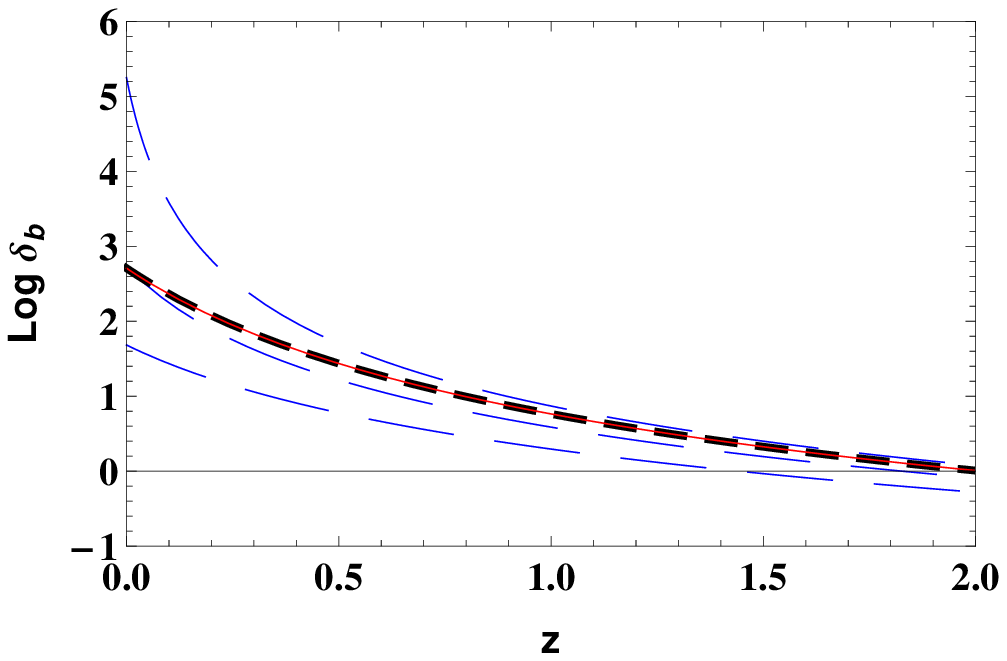}
%\label{fig1}	
\caption{Perturbation growth of baryons as a function of the redshift. The same as Fig.2 but now we add the baryonic perturbation in the bulk viscous fluid model (dashed blue lines) with, from top to bottom, $\tilde{\xi}=0.3, 0.5$ and $0.7$ fixing $\nu=1/4$.  }
\end{center}
\end{figure}

\begin{figure}\label{fig10}
\begin{center}
\includegraphics[width=0.43\textwidth]{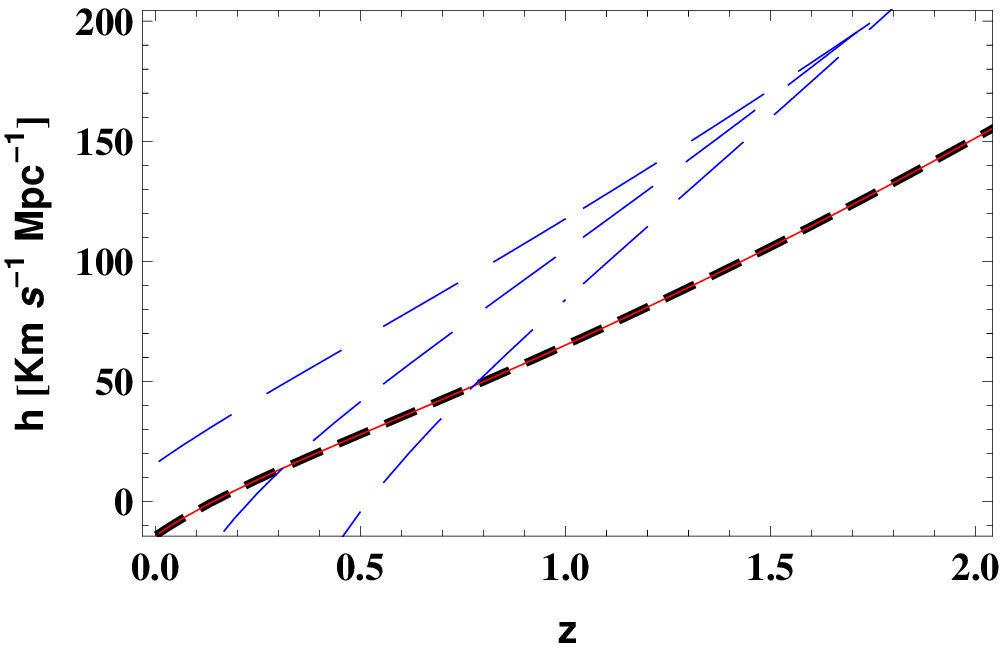}
%\label{fig1}	
\caption{Evolution of the expansion rate of the collapsed region. The same as Fig.3 but now we add the result for the bulk viscous fluid (dashed blue lines) with, from top to bottom, $\tilde{\xi}=0.3, 0.5$ and $0.7$ fixing $\nu=1/4$. }
\end{center}
\end{figure}

\begin{figure}\label{fig11}
\begin{center}
\includegraphics[width=0.43\textwidth]{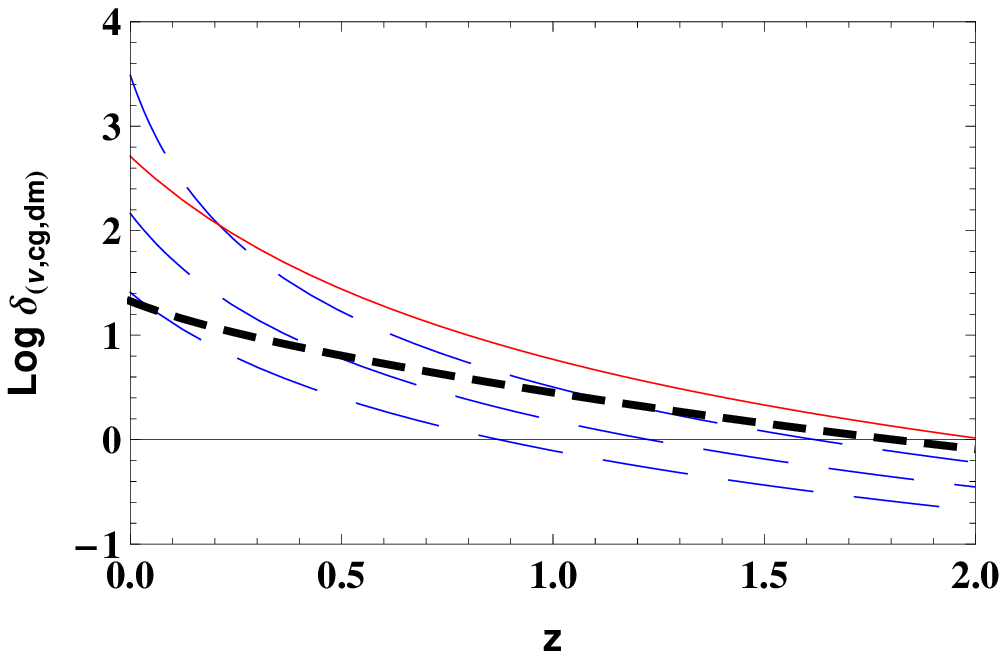}
%\label{fig1}	
\caption{Dark matter perturbation growth as function of the redshift. The same as Fig.1 but now we add the perturbation in the bulk viscous fluid $\delta_v$ (dashed blue lines) with, from top to bottom, $\tilde{\xi}=0.05, 0.075$ and $0.1$ fixing $\nu=1/2$. }
\end{center}
\end{figure}

\begin{figure}\label{fig12}
\begin{center}
\includegraphics[width=0.43\textwidth]{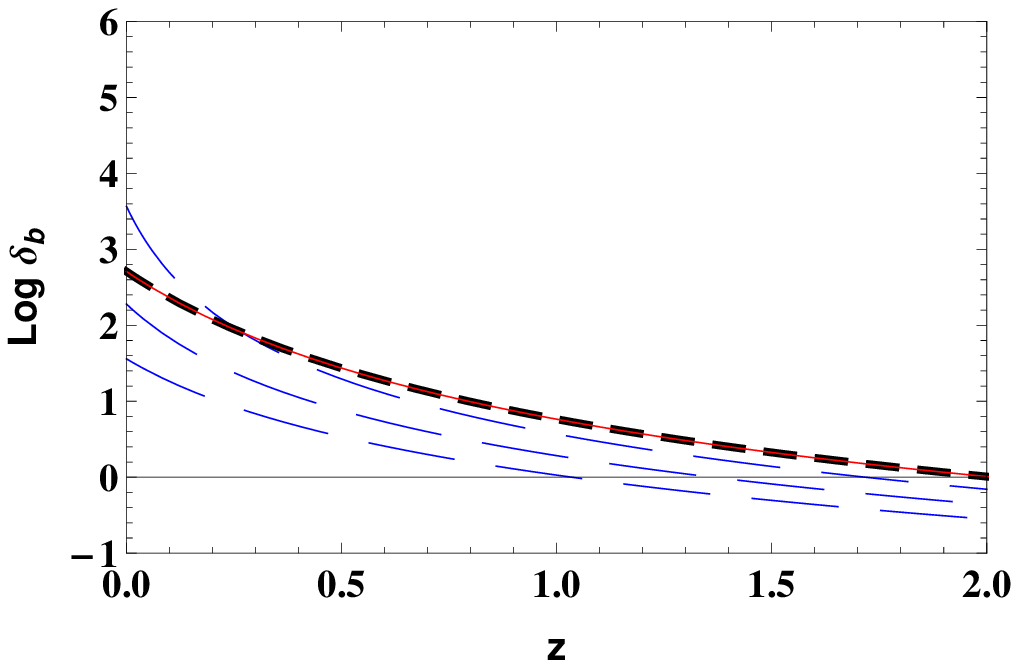}
%\label{fig1}	
\caption{Perturbation growth of baryons as a function of the redshift. The same as Fig.2 but now we add the baryonic perturbation in the bulk viscous fluid model (dashed blue lines) with, from top to bottom, $\tilde{\xi}=0.05, 0.075$ and $0.1$ fixing $\nu=1/2$.}
\end{center}
\end{figure}

\begin{figure}\label{fig9}
\begin{center}
\includegraphics[width=0.43\textwidth]{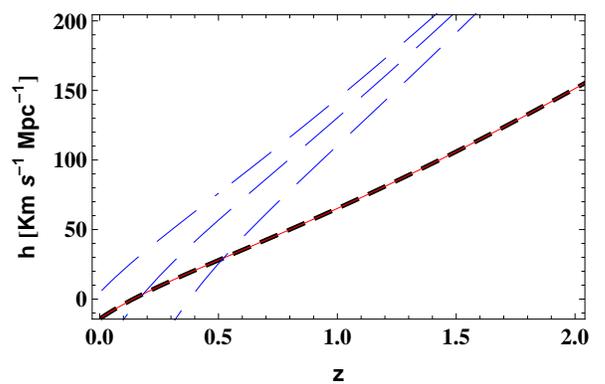}
%\label{fig1}	
\caption{Evolution of the expansion rate of the collapsed region. The same as Fig.3 but now we add the result for the bulk viscous fluid (dashed blue lines) with, from top to bottom, $\tilde{\xi}=0.05, 0.075$ and $0.1$ fixing $\nu=1/2$. }
\end{center}
\end{figure}

\section{Conclusions}

We have studied the spherical collapse for the class of unified cosmologies. We worked with 
two candidates namely, i) the generalized Chaplygin gas and ii) the bulk viscous model. For each model we had distinct aims. 

\begin{itemize}

\item For the GCG our goal was to understand its $\Lambda$CDM limit ($\alpha=0$) and then 
promove a suitable comparison with the standard cosmology in order to test the so-called ``dark degeneracy''. 

The comparison between the GCG with $\alpha=0$ and the $\Lambda$CDM shows that although the visible dynamics, i.e. the baryonic part and the expansion of the collapsed region, is indeed the same, the dynamics of the dark part of such models is different. This result supports the conclusion that the models are not exactly the same because they would display different bias factors. The relation between the baryonic and the dark matter overdensities is different. This issue is the key point for understanding the so-called ``dark degeneracy'' \cite{Kunz}. Therefore, lensing observations could be able to differentiate such cosmologies.

\item For the bulk viscous model we employed for the first time a non-linear study and our main goal was 
to understand the magnitude of the viscous parameter that allows structure formation at non-linear level.

The bulk viscous model does lead to virialized structures, but only for $\tilde{\xi}$ values that do not provide the large scale background expansion of the universe. Therefore, this result challenges unified viscous models since collapsed dark viscous structures would never exist in the unified scenario. We have checked the consistence of this result for different values of the parameter $\nu$ which shapes the form of the bulk viscous coefficient $\xi$. In some sense the viscous effects prevents structure formation at non-linear order because the coefficient $\tilde{\xi}$ assumes high values. In the unified picture this is a necessary condition because of the need to accelerate the background expansion. An interesting scenario for the bulk viscous dark matter occurs in the context of the standard cosmology when only the cold dark matter presents dissipation while the cosmological constant accelerates the universe. This is the essence of the $\Lambda$vCDM model \cite{dominikVelten2012, velten, velten2014}. The suppression mechanism present in the bulk viscous dark matter seems to be an indication for solving the small scale problems of the standard model which are related to the excess of satelites and the cusp profile of galaxies \cite{velten2014}. In a future communication we will study the non-linear aspects of the $\Lambda$CDM model. 

\end{itemize}

\textbf{Acknowledgement}:  We thank CNPq (Brazil) and FAPES (Brazil) for partial financial support.

\end{document}